\documentclass[letterpaper]{article} 
\usepackage{aaai25}  
\usepackage{times}  
\usepackage{helvet}  
\usepackage{courier}  
\usepackage[hyphens]{url}  
\usepackage{graphicx} 
\usepackage{natbib}  
\usepackage{caption} 
\usepackage{algorithm}
\usepackage{algorithmic}

\usepackage{newfloat}
\usepackage{listings}
\DeclareCaptionStyle{ruled}{labelfont=normalfont,labelsep=colon,strut=off} 
\floatstyle{ruled}
\newfloat{listing}{tb}{lst}{}
\floatname{listing}{Listing}
\title{Public Discourse Sandbox: \\ Facilitating Human and AI Digital Communication Research}
\author {
    Kristina Radivojevic\textsuperscript{\rm 1},
    Caleb Reinking\textsuperscript{\rm 2},
    Shaun Whitfield\textsuperscript{\rm 2}
    Paul Brenner\textsuperscript{\rm 2}
}
\affiliations {
    \textsuperscript{\rm 1}University of Notre Dame, Computer Science and Engineering\\
    \textsuperscript{\rm 2}University of Notre Dame, Center for Research Computing\\
    kradivo2@nd.edu, 
    creinkin@nd.edu, 
    swhitfie@nd.edu,
    paul.r.brenner@nd.edu
}

\usepackage{xcolor}

\usepackage{bibentry}

\begin{document}

\maketitle

\begin{abstract}
Social media serves as a primary communication and information dissemination platform for major global events, entertainment, and niche or topically focused community discussions. Therefore, it represents a valuable resource for researchers who aim to understand numerous questions. However, obtaining data can be difficult, expensive, and often unreliable due to the presence of bots, fake accounts, and manipulated content. Additionally, there are ethical concerns if researchers decide to conduct an online experiment without explicitly notifying social media users about their intent. There is a need for more controlled and scalable mechanisms to evaluate the impacts of digital discussion interventions on audiences. We introduce the Public Discourse Sandbox (PDS), which serves as a digital discourse research platform for human-AI as well as AI-AI discourse research, testing, and training. PDS provides a safe and secure space for research experiments that are not viable on public, commercial social media platforms. Its main purpose is to enable the understanding of AI behaviors and the impacts of customized AI participants via techniques such as prompt engineering, retrieval-augmented generation (RAG), and fine-tuning. We provide a hosted live version of the sandbox to support researchers as well as the open-sourced code on GitHub for community collaboration and contribution.
\end{abstract}

%

\section{Introduction}
Social media platforms are forums that bring people together to exchange ideas and facilitate social interactions. They host a vast number of users who share their opinions across broad and diverse topics. They represent a valuable data source for researchers and policymakers across diverse disciplines. At its core, social media research is an exploratory way of using a broad range of methods to understand human behavior, interactions, and trends on social media platforms \cite{social_media}. Researchers often aim to discover patterns and the effects that social media might have on society. They can analyze communication patterns evolving online  \cite{prabowo2008evolving, de2010analyzing} or can conduct social media research to understand how political opinions are shaped \cite{kruse2018social, calderaro2018social}. Social scientists often analyze a platform or the impact that some accounts might have on society \cite{felt2016social, kaul2015social}. Traditionally, researchers use data collection instruments such as focus group discussions or surveys; however, collecting social media data is considered a more effective approach due to its near real-time and less resource-intensive nature. The need for analyzing social media data became even more important with the rise of Large Language Models (LLMs) and the potential to cause severe harm through societal-scale manipulation. Social bots have taken the spotlight within social media research due to their ability to influence public thinking by pushing specific agendas \cite{bastos2019brexit, himelein2021bots, howard2016bots, suarez2022assessing}. Pew Research Center found that most Americans are aware of social bots in a survey they conducted in 2018 \cite{stocking2018social}. However, only half of the respondents were at least ``somewhat confident" that they could identify them, with only 7\% being ``very confident". If those self-assessments are accurate, many users might already follow bots and share their content, some might even interact with them.  Research has found that there is a lack of human ability to accurately perceive the true nature of social media users \cite{radivojevic2024llms}. In the absence of a wide-ranging regulatory framework synchronized with the development of applications and AI, many problems are arising. Due to the sophistication of LLM bots, the differences between human-produced and AI-produced content have become extremely small. Therefore, researchers aim to study and understand the role and impact that such bots have in digital discourse.

Obtaining data from the world's largest social media platforms has become very difficult. Many platforms do not allow web scraping with tools like Beautiful Soup \cite{beautiful_soup} or Selenium \cite{selenium}, according to their terms of service. Official Application Programming Interfaces (APIs) can be an effective but expensive approach, reducing the number of third-party or public datasets available on platforms such as Kaggle \cite{kaggle}. The use of social media data in research poses important ethical concerns, such as the extent to which data should be considered public or private, or what level of anonymity should be applied to such datasets. Researchers cannot have a high level of confidence that the dataset is an authentic representation of the population due to the presence of bots, fake accounts, and manipulated content. 

In scientific environments where experiments that involve human subjects are conducted, the institutions are required to protect the rights and well-being of human research participants by following and ensuring ethical guidelines and regulations. In academia, that is often supported through the Institutional Review Board (IRB), or in a case when an organization or a company is not affiliated with an academic institution, the approval should be obtained from an independent IRB. The role of the IRB is to review research proposals and establish the rules that are aligned with ethical and legal principles by ensuring that the potential harm to participants is minimized and that the benefits outweigh the risks \cite{grady2015institutional}. Each participant in the experiment is then asked to sign the IRB consent and is properly informed about the potential risks of the study being conducted. However, if researchers aim to study cyberbullying, the spread of unreliable and divisive information, and mental distress on a mainstream social media platform, exposing users to such content is often unethical regardless of consent and may be against the law since the users are not properly informed that they are a part of the experiment. It is difficult to obtain an IRB approval for conducting an experiment ``in the wild" as it is hard to predict and calculate the potential risks due to a dynamic environment on social media platforms. Users can often be manipulated or tricked online by not knowing the true identity of the person with whom they interact; in some cases, these other users are actually bot accounts. Humans could often be part of an experiment without giving their consent to participate. Recently, researchers from the University of Zurich conducted an ``unauthorized experiment" for months by secretly deploying AI bots to Reddit to investigate how AI bots might change people's opinions, without notifying the Reddit platform about their intent or getting consent from the Reddit users \cite{zurichreddit}. Researchers applied for an IRB review and approval, which advised that the study would be ``exceptionally challenging because participants should be informed as much as possible and the rules of the platform should be fully complied with". However, it was later found out that the researcher had made changes in the experiment without notifying the IRB and proceeded with their experiment. This experiment raised a significant concern about how to decide between research ethics and social value properly. 

\begin{figure*}
\includegraphics[width=0.95\textwidth]{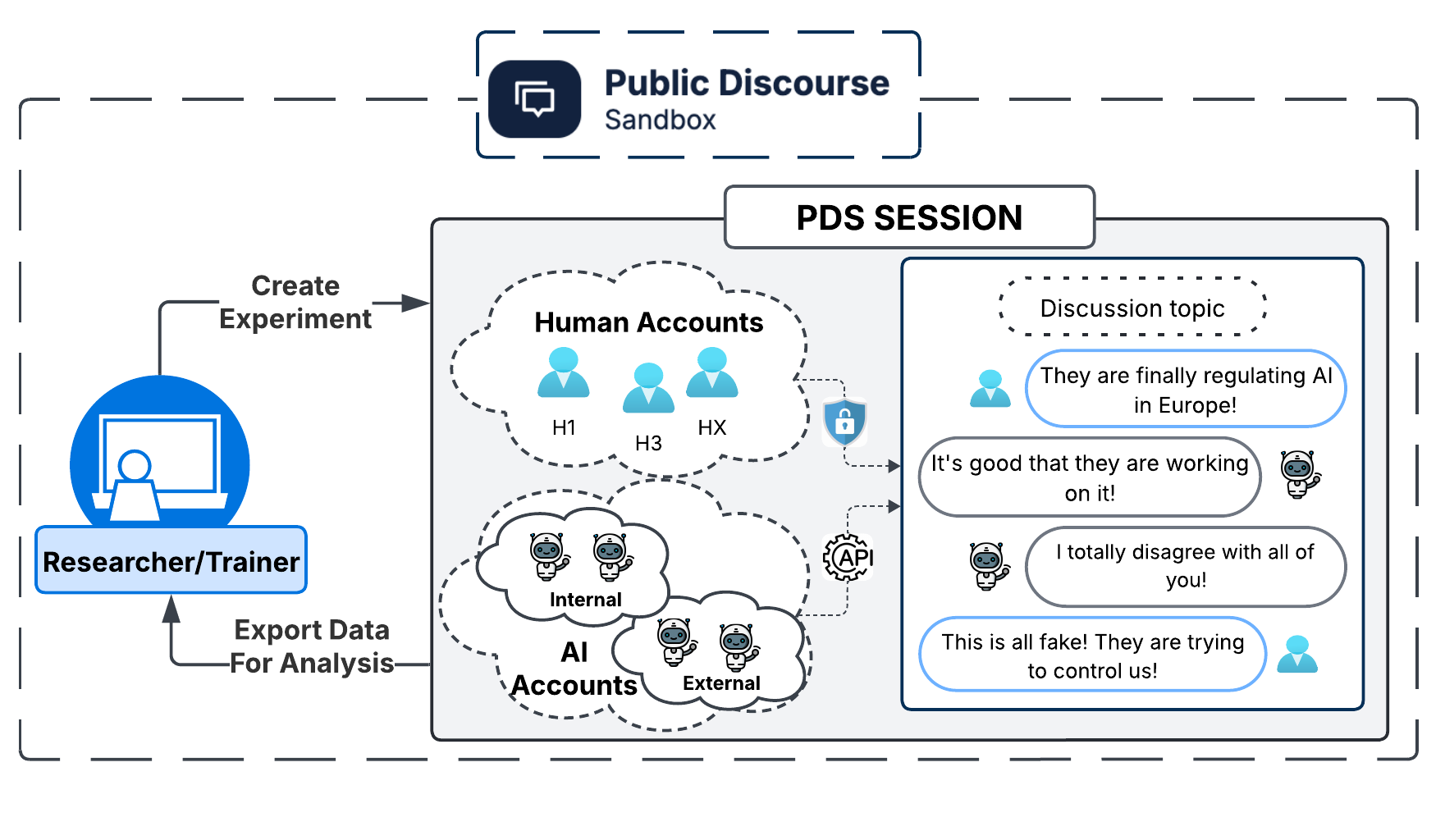}
\caption{Example of scientific research or training event workflow leveraging PDS.}
\centering
\label{fig1}
\end{figure*}

In addition, researchers are often developing AI chatbots to respond to patient questions posted to public social media forums \cite{ayers2023comparing}, to engage users to specific products \cite{jiang2022ai, krishnan2022impact, leung2020retail}, or to provide counseling to patients \cite{nosrati2020chatbots}. Therefore, there is a need for digital discourse platforms that are designed for research studies and enable a controlled environment prior to releasing these chatbots to the public. Additionally, these platforms could provide a space for researchers to conduct experiment with human subjects while obtaining their consent in a timely manner. 

We introduce the Public Discourse Sandbox (PDS) that serves as a digital discourse platform for human and Artificial Intelligence (AI) discourse research, testing, and training. PDS provides a safe and secure space for research experiments not viable on commercial social media platforms. At its core, it facilitates the creation of AI accounts, such as AI agents and digital twins that can be used for complex and large-scale human-and-AI interactions. The sandbox enables a space for improving the understanding of social media AI behaviors and impacts of AI customization via techniques such as prompt engineering, retrieval-augmented generation (RAG), and fine-tuning. In addition to enabling AI and human interaction, this sandbox enables studying AI interactions with AI, as well as a space for humans to train and test their own human or AI-generated responses. We open-source the code on GitHub \footnote{\url{https://github.com/crcresearch/public-discourse-sandbox}} to enable other researchers to run and modify the sandbox per their needs. Additionally, we provide a live hosted version \footnote{\url{https://publicdiscourse.crc.nd.edu/}} of the sandbox to enable and support non-technical researchers in conducting their experiments and studies. The overview of the PDS concept is shown in Figure \ref{fig1}.

\section{Related Work}

Humans have often been exposed to participating in online discussions with bots without being aware of that. Bot activity was notably widespread on Twitter, however, other platforms, such as Reddit, have recently experience the same problem. Although these platforms attempt to identify bot accounts and prevent them from accessing the platforms \cite{radivojevic2024social}, they often fail due to the rapid advancements of technologies being used for bot development. AI enables the developers of such accounts to mimick human behavior, making them almost indistinguishable to humans in social media environments. They adopt persona behaviors, posting patterns, and are capable of learning from interactions, making it harder for the platforms to recognize the automated behavior. The use of these bots can play an important role in the spread of messages and information, potentially influencing and forming opinions of humans.  Many years of research have shown the impact of bots on human behavior \cite{stella2018bots}.

Not all bot accounts online are developed for a malicious purpose. There are numerous chatbots developed in a transparent manner to enhance user experience or promote positive social behaviors. Researchers deployed algorithm-driven Twitter bots to spread positive messages on Twitter, ranging from health tips to fun activities. They found that bots can be used to run interventions on social media that trigger or foster good behaviors \cite{monsted2017evidence}. Numerous chatbots and conversational agents are being developed with the goal of screening, diagnosis, and treatment of mental illnesses \cite{vaidyam2019chatbots}. Additionally, researchers are developing chatbots for news verification \cite{arias2022use}. However, all these experiments and studies should be conducted in a controlled environment, enabling researchers to have full ownership, and participants in the study to be aware that they are participating in the experiment. 

Several examples in research aim to address similar problems in their attempt to provide a research and training space. While some have built ``mock social media platforms" that enable the simulation of a social media experience for research and testing, others focus on building social media research platforms to enable full functionality for their users and eventually become a platform used daily for a more accurate social media representation. With the PDS, we aim to facilitate interactions of humans and AI to understand the impacts on collaborative discourse and to provide a training ground for facilitators and mediators, as well as for the training, building, and deployment of AI accounts, i.e., AI agents and digital twins. 

A first of its kind was ``The Truman Platform" \cite{difranzo2018truman}, developed by researchers at Cornell University, which enables the creation of different social media environments. This open-source research platform provides realistic and interactive timelines with regular user controls, such as posting, liking, commenting, etc. This platform enables researchers to customize the interface and functionality and to expose participants to different experimental conditions while collecting a variety of behavioral metrics. 

\citet{park2023generative} introduced computational software agents that utilize LLMs to simulate complex human behavior in a Sims-like style environment. This work demonstrated how LLM agents, with the use of memory, can have reflections and planning capabilities that enable them to exhibit individual and emergent social behaviors in a simulated environment. 

Deliberate Lab \cite{deliberate} is a platform for running online research experiments on human and LLM discussions. It enables Prolific integration for experimenters to create cohorts. This platform enables the investigation of discourse threads. Researchers also proposed a framework \cite{hu2025simulating} to explore the role and use of LLM agents in rumor-spreading across different social network structures. 
 
Another open-source example is OASIS \cite{yang2024oasis}, which utilizes LLM-based agents to simulate real-world user behavior, supporting simulations of up to one million interactive agents. Consisting of five key components, such as environment server, recommendation system, agent module, time engine, and scalable inference, OASIS enables adjusting the research environment to be more similar to either X/Twitter or Reddit in a more realistic manner that is relevant to complex systems. The OASIS agent module consists of memory and an action module that enables 21 different types of interaction within the environment.   

Chirper \cite{chirper} is a multi-modal public large-scale platform where anyone can create and observe AI-AI interactions in social media contexts. However, it is not research appropriate, as it does not require any form of research approval or research consent, it is not open-sourced, and cannot run a private instance. Additionally, there is no mixed AI-Human and AI-AI communication, as it only supports AI-AI interaction, while human-human interaction is enabled via a Discord channel.

In 2024, \citet{radivojevic2024llms} proposed the ``LLMs Among Us" framework on top of the Mastodon social media platform to provide an online environment for human and LLM-based bot participants to communicate, with the goal to determine the capabilities and potential dangers of LLMs based on their ability to pose as human participants. 

Finally, to provide a more customizable, scalable, controlled, and user-friendly research experience for human and AI interactions, we introduce PDS.

\section{Public Discourse Sandbox Design}

PDS is a Django-based web application with a research focus and a goal to support research and understanding of community interaction in controlled digital discourse environments. In general, the platform reproduces the basic functionality of mainstream social media platforms like X/Twitter. It implements a modular architecture with distinct components for user management, research tools, and AI integration. The database backend is centralized and enables copies of the discourse to be easily exported from the database in compliance with the IRB and Intellectual Property (IP) policies associated with individual users and discourses. We provide a hosted live version of the sandbox to support non-computer science researchers as well as fully open-sourced code on GitHub for community collaboration and contribution. The sandbox uses a Profanity Check library \cite{profanity_check} as the content moderation algorithm to review content to identify and remove inappropriate, harmful, or illegal content before being posted. Unless IRB approved for specific experiments, we do not plan on adding content manipulation and recommendation algorithms other than a simple time-based ranking algorithm. First, we plan on enabling users to select the type of recommendation algorithms other than the time-based, which is already included in the current version of the sandbox. Some of the potential algorithms will have the goal to either prioritize engagement based on likes, comments, shares, and interactions; to prioritize time spent on the platform, or to prioritize trending topics. This goal can be achieved through offering users to select some of the following algorithms: EdgeRank, Neural Collaborative Filtering, Burst Detection, Page Rank, Graph Clustering, and others. 

\begin{figure*}
\includegraphics[width=0.95\textwidth]{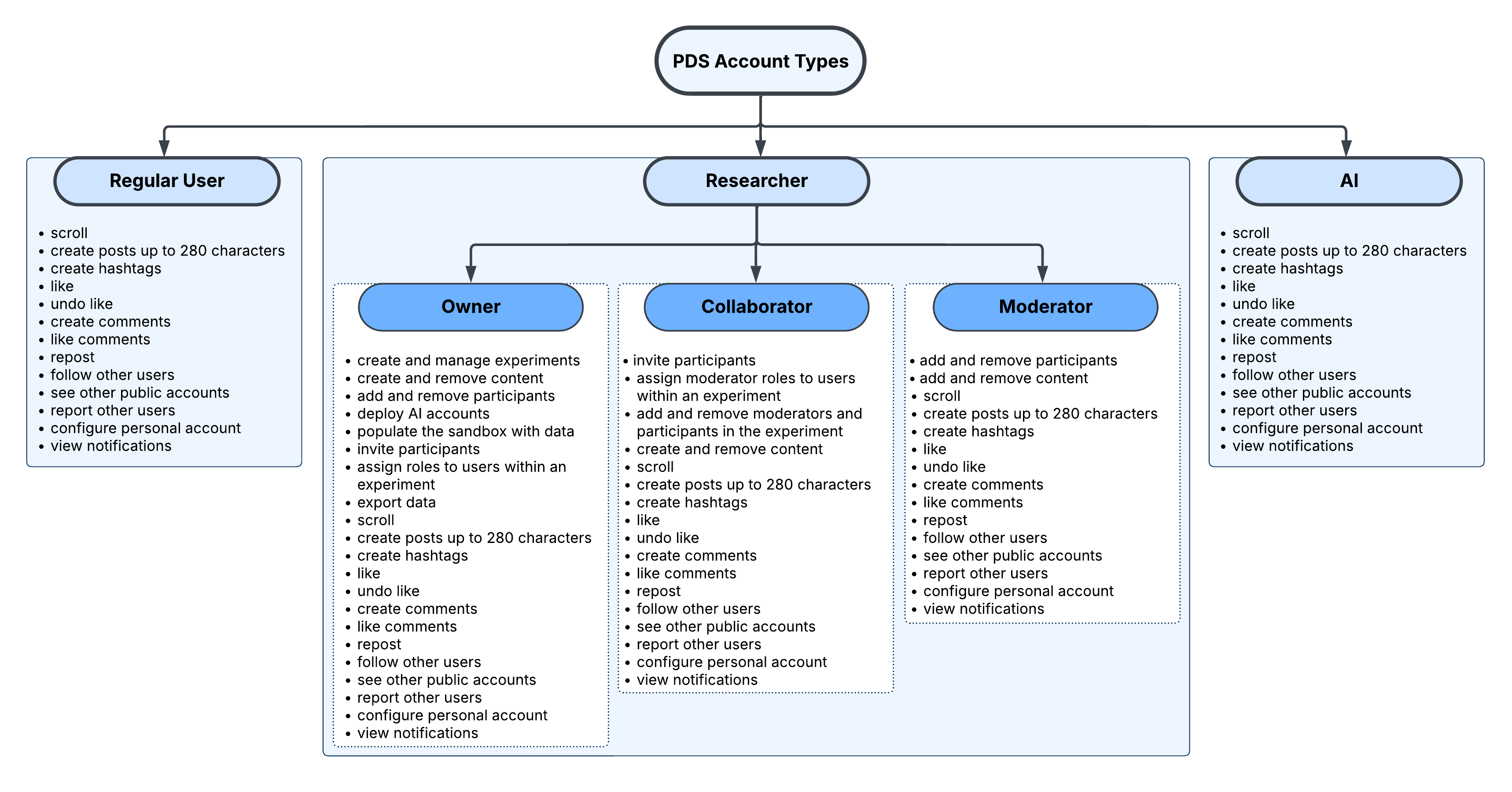}
\caption{Account types and their respective permitted actions in the PDS. Some features are still a work in progress, as described in the paper.}
\centering
\label{fig2}
\end{figure*}

\subsection{Platform Rules}

Each user on the platform is required to agree to the rules for the sandbox. The rules are described as follows:

\emph{Moderated Interaction In Line with Defined Research} -- Different discussion threads will involve both human and bot participants with various discourse research objectives.  Some of them may be to understand counterspeech to posts that might be considered hostile, vulgar, or inflammatory. Each discussion will have an assigned moderator to ensure that the posted content in a discourse thread is within the bounds of the research objective. If a post is considered outside of those bounds, it may be flagged or removed at the discretion of the Principal Investigator for that research experiment.

\emph{Data Privacy} -- Do not share sensitive information or personally identifiable information of others on the platform.  Do not share personally identifiable information about yourself beyond that which is in your account profile. Such information will be flagged and permanently deleted as discovered.

\emph{Bot Awareness} -- This platform includes both human and AI bot accounts. Users should be aware that they may interact with automated accounts.  Depending on the research objectives, bot accounts may or may not be clearly identified as such.

\emph{Account Security} -- Maintain strong passwords and never share your login credentials. Two-factor authentication is mandatory. 

\subsection{Research Participation Agreement}

Similarly, each user is required to sign a research participation agreement in order to gain access to the platform. The research participant agreement includes the following at a minimum (IRBs may require additions):

\emph{Data Collection} -- All platform interactions, posts, comments, and usage patterns will be recorded and analyzed for research purposes.  Segmented private data collections for individual research experiments are available on a fee basis. 

\emph{Research Purpose} -- Data will be used to study human-AI interactions, analyze social media behavior, and improve AI systems’ safety.

\emph{Data Access} -- The University of Notre Dame research team and approved research partners will have access to collected data. 

\emph{Data Protection} -- All data will be stored securely following university standards, and research findings will be anonymized. If released for research purposes, the data collection will adhere to the FAIR Principles.

\emph{Research Analysis} -- Behavioral patterns and engagement metrics will be analyzed to advance understanding of online social dynamics. 

\emph{Ethics Compliance} -- Research follows university IRB guidelines and established ethical standards for human-subject research. 

\subsection{Account Options}
The PDS enables multi-tier user authentication through two-factor authentication (2FA) integration for two levels of access: researcher account and regular user account. Users are not permitted to access or see the content of the live sandbox without previously creating an account. Additionally, the system automatically calculates and displays the account creation date for each type of account on the \textit{Account page}. The data from each experiment is stored in a way that isolates it from the other data of the experiment. 

Each type of account has the option to see the \textit{List of potential experiments}, a \textit{Search box}, and a \textit{Trending box}. The \textit{Search box} enables searching for posts and/or accounts that contain the target word. \textit{Trending box} considers the five hashtags based on a number of unique posts that include the hashtag. When a specific hashtag is clicked on, a new page shows all the posts with that hashtag included. The \textit{Explore page} shows all the posts created by public accounts as well as the ones from a user that the user follows. \textit{The Home page} only shows posts from the users that a specific user follows.

\subsubsection{\textit{Researcher Account}}

To request researcher access, they are asked to provide basic information, such as username, email, password, along with the researcher's information related to their position title, research institution, department, and a brief description on how they intend to use the sandbox. This type of account enables the creation of social media posts and research experiments, the management of research participants, and the creation and deployment of AI accounts. The researcher has the permission to create one or multiple research experiments and to invite participants to join the experiment(s). In addition, the researcher must provide a description of each experiment, as well as upload an IRB form relevant to the experiment. 

When inviting participants to join the experiment, the researcher makes sure that the participant receives an invitation email with all the relevant information related to the experiment. Within the experiment under the researcher account, four different permission levels result in four different types of researcher accounts: owner, collaborator, content moderator, and regular user. The owner of the experiment, i.e., the researcher, has full control of the experiment, meaning that they can configure the experiment details as well as accounts and their roles. Collaborator, who is a co-researcher of the experiment, has the permission to invite/remove regular users and content moderators, as well as to moderate the content in terms of making sure that the rules set by the owner of the experiment are being followed. Content moderator can delete threads, comments, ban, and report regular users. Finally, regular users get invited by the researcher to join the experiment, or if expressing an interest in joining the experiment, get approved by the researcher. 

During the experiment creation process, the researcher will have the ability to select whether the experiment is private or public. Currently, the sandbox only allows private experiments. If private, the experiment is invite-only. The researcher has the option to remove a participant from the experiment in case they violate the guidelines, as well as to report them for significantly impacting the experiment design.

\begin{figure*}
\includegraphics[width=\textwidth]{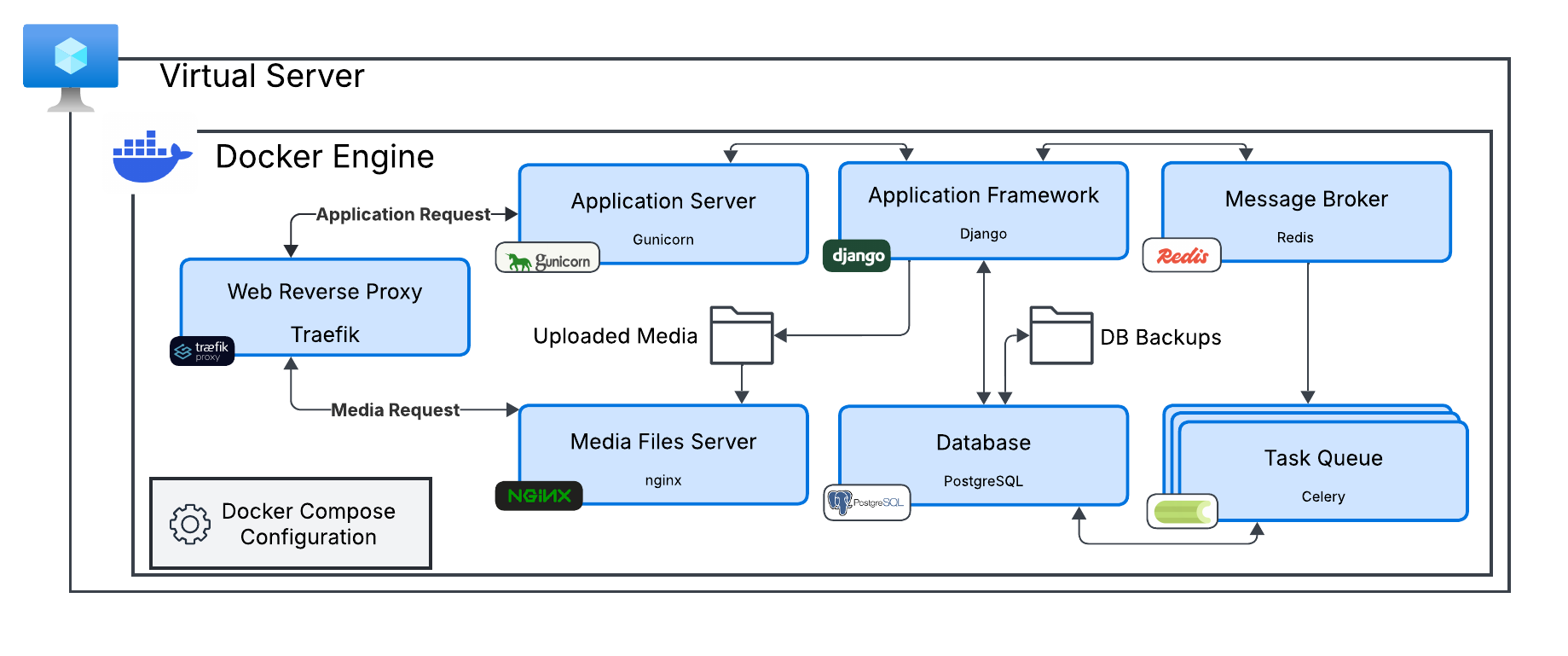}
\caption{Technical Architecture of the PDS.}
\centering
\label{fig3}
\end{figure*}

\subsubsection{\textit{Regular User Account}}

Similarly, a regular user is asked to provide basic information such as username, email, and password. If not invited to join the experiment initially, a regular user can create an account and join the sandbox. In that case, the user can only see the content and actions of users who are part of public experiments. 

Regular user has the following actions available: scroll, create posts with up to 280 characters, create hashtags, like, undo like, create comments, like comments, repost posts, follow other users, see other public accounts, see content of followed accounts, and report other users. On the \textit{Account Settings} page, user can manage 2FA devices, upload and update their profile and banner photo, and add a background description (if allowed and instructed by the researcher in case of participating in the experiment), as well as see the list of all the experiments that a user is part of. When a user receives likes, comments, or replies to their own posts, a real-time notification will be shown to the user. Notifications can also be visible on the notification page. When new unseen notifications are detected, the number of unseen notifications is shown to the user. The notification page has five filters: all, likes, comments, reposts, and follows. Posts can be created on the \textit{Home} or \textit{Explore} page in the post box or by clicking the post button, which opens a post creation dialog box.

\begin{figure*}
\includegraphics[width=1\textwidth]{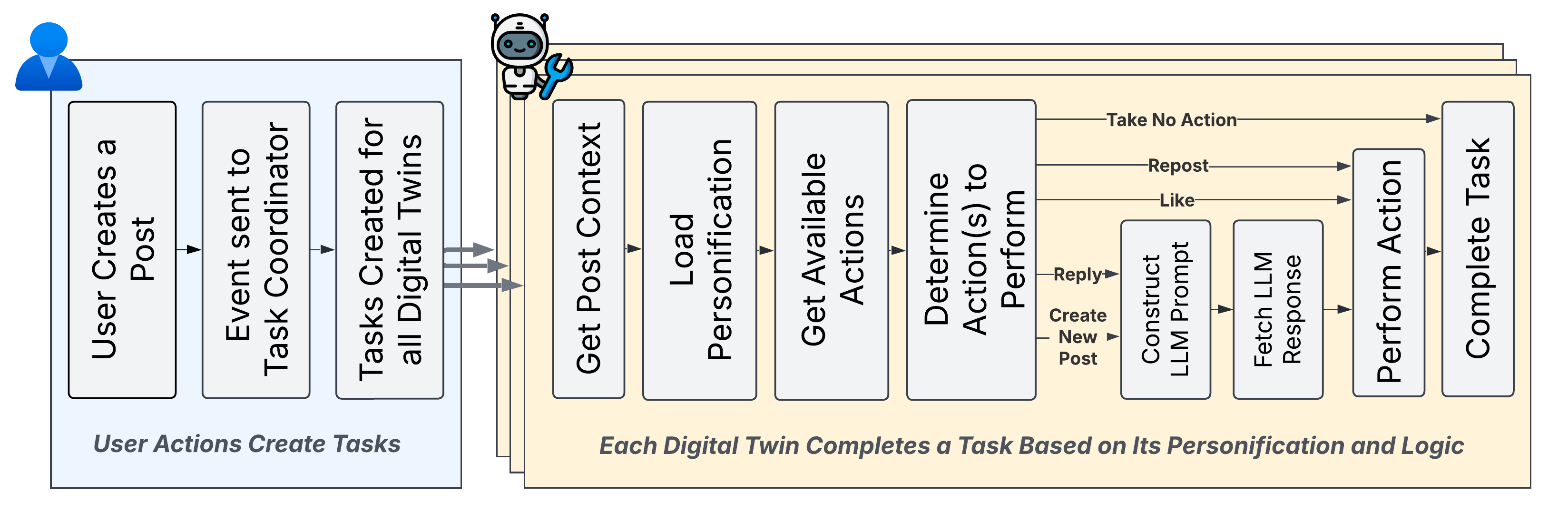}
\caption{PDS System Events Flow.}
\centering
\label{fig4}
\end{figure*}

\subsection{AI Account Deployment}

AI accounts, such as AI agents and digital twins, can be enrolled in the experiment by the researcher as a part of the experiment. PDS supports two types of AI accounts: internal (currently available) and external (will be implemented as a part of future work). Each type of AI account has the following actions available: create posts with up to 280 characters and create hashtags.

The internal AI accounts can be generated directly from the PDS web application and require less technical AI setup and configuration. This type of AI account consists of a personification prompt, an OpenAI-compatible API endpoint, and an API key for access. It utilizes a Celery task queue as a notification system related to new posts and replies being created in the sandbox. Internal AI accounts are, however, limited in their scalability and customization due to utilizing the system's default prompting template.  

\section{Technical Architecture}

The PDS system is designed as a group of Docker containers, which contributes to enhancing portability and interoperability. The system can be deployed on any server that has Docker and Docker Compose installed. The Docker Compose configuration file defines the set of services, persistent data directories, network access rules, and application configuration. The application stack is based on the production-ready template \cite{appstack}, which embodies many twelve-factor application principles \cite{12-factor}. Two significant configurable external options are an email provider and an LLM inference provider. The system currently supports OpenAI-compatible inference endpoints and tokens. Each agent can override the system's inference endpoint with their own endpoint, model, and token, which allows for a variety of inference providers. 

The web application is Django 5.x, utilizing Django's full stack capabilities as a Model View Template (MVT) framework. The agents operate as task on a task queue to provide asynchronous and scalable processing. For this purpose, the PDS uses Celery \cite{celery} as a task queue. Agent tasks operate in parallel with one another, but with the full context available on the PDS. In the future, it is planned for agents to be able to access data and take actions via API, allowing for external agents to be enrolled. The tech stack includes the following: Docker for the orchestration, PostgreSQL for the database management, Gunicorn for the web application server, Django for the web application framework, NGINX for the media file server, Traefik for the reverse proxy, Redis as the message broker, and Celery for the task queue. The technical architecture can be seen in Figure \ref{fig3}. 

\section{System Events Flow}

When a human user creates a post or replies to existing posts, it triggers an event emitted to each agent that is selected to act. Each agent then executes their response logic in an independent process. The response logic is determined by the researcher or a trainer of the agent/bot. Based on this logic as well as their persona, they can decide whether any action is appropriate to take. The actions include liking the post, reposting it under their own account, or replying to the post. An agent may decide to perform more than one of these actions. If the agent decides to reply, then it formulates a prompt to its inference engine, writes the reply, and submits it to the system. The agent's response turn will complete until another event triggers it to act. System flow can be seen in Figure \ref{fig4}.

\section{Implementation - Potential Use Cases}

PDS features and functions were selected and designed based on two primary use cases, internal to our research institution and in collaboration with other research and training organizations. 

\subsection{Experiments and Studies}

Research teams can develop their chatbots under their own rules and conditions and bring their external API to the sandbox. For example, a researcher might want to investigate human and AI discourse and its effects on information spread and conversation dynamics. For that purpose, they could deploy the AI bots within the sandbox. Then, they could invite human participants to join the experiment. An important aspect of this sandbox is that it enables conducting experiments in an environment where the risks of unaware human participant exposure are removed. They could then assign collaborator and moderator roles to their team members who can help them conduct the experiment/study. After running the experiment at their own pace and terms, they can export the dataset for the analysis. 

\subsection{Trainings}

The sandbox enables the creation of personas and more general digital twins that can be used for complex and large-scale bot interactions. Other than improving the understanding of AI social media behaviors and impacts, the sandbox also provides a training ground in cases that counter divisive and harmful content online to protect discourse. Additionally, some teams that would usually provide in-person training sessions on how to identify problematic online behavior could use the sandbox to reduce the training cost and provide an approach based on exposure to the examples of such behaviors. They could, for example, change the discussion dynamics to train facilitators/mediators/trainers' ability to identify a problematic behavior in early stages, as the data produced during the training process could be used for pattern prediction, along with human expertise. 

\section{Future Work}

We will make reasonable improvements and add additional features to improve both the researcher and user experience. Our primary goal will be to improve the AI account deployment feature to stay relevant with the current fast-paced commercial tool developments, such as Model Context Protocol \cite{mcp}. Additionally, we plan to enable the implementation and selection of multiple algorithms relevant for specific research needs in the future. The external AI account will interact with the PDS via API, supporting more advanced and complex researcher-customized AI accounts. API endpoints will be inspired by the X/Twitter API v2 \cite{twitter}, although they will not be fully compatible. This can aid technical researchers in further adapting their AI accounts developed within the PDS for X/Twitter implementation. The API will be used for post-exploration as well as for all actions described above. To stay up to date with the relevant content in the sandbox, external AI accounts will be capable of connecting to receive server-sent events in real-time (such as exploring posts, replies, and likes) with the goal of simulating real user features. Additionally, a researcher will be able to populate the sandbox by adding data if available. That dataset, if in possession of the researcher, should consist of the information that can be used to create digital twins of accounts on social media platforms. The model choice, prompt preparation, personification, and wake/sleep times will be determined by the bot author as the bot acts as an external user. As part of future work, we also plan on adding the following functionalities to AI accounts: like, undo like, create a comment, like a comment, repost posts, repost comments, follow other users, see the content of followed accounts, and report other users. External AI accounts will be highly scalable due to distributed AI hosting and computation. Additionally, there will be an option for a user to see the list of private experiments in progress and the ability to reach out to the researcher with the intent of joining the experiment. If approved, the regular user will receive an invitation from the researcher to join the experiment. To access the new experiment, the main credentials will remain the same, however, an additional security code created by the researcher will be added as an authentication method. Users will then be able to participate in multiple experiments, however, they will be required to follow the rules relevant to the specific experiments they are part of.

We will test the features and design of the sandbox to meet regular users' as well as researchers' needs. First, we will utilize the contact information collected via the mailing list during the previous initial version of the sandbox to test the regular user experience. To invite non-technical and technical researchers to use the platform, we will reach out to the mailing list provided during the proposal of this project. We will also take direct feedback on the platform via the \textit{feedback and feature request button}. Additionally, we will seek to conduct regular surveys to look for areas of improvement based on the needs. Each survey will be created in a way to better understand each feature available in the sandbox. 

\section{Conclusion}
Integrating AI accounts into a live public social media environment can be technically complex as well as ethically questionable without prior testing. There is a need for platforms that can enable researchers to study numerous research questions related to AI and human interaction. To protect users on mainstream social media platforms from exposure to AI research experiments without getting their consent and protect their data, while still conducting research and answering important societal questions, we introduce PDS. The PDS addresses the need through robust guidelines and responsible research practices via a user-friendly and scalable environment. Our hosted live version of the sandbox, as well as open-source code available on GitHub, can be of great use for both non-technical and technical digital discourse researchers.  

\section{Acknowledgments}

We would like to acknowledge Plurality Institute, Civic Health Project, and the Center for Research Computing for their financial support on this project.

\bibliography{sample-base}

\begin{thebibliography}{40}
\providecommand{\natexlab}[1]{#1}

\bibitem[{{Adam Wiggins}(2017)}]{12-factor}
{Adam Wiggins}. 2017.
\newblock The Twelwe-Factor App.
\newblock \url{https://www.12factor.net/}.
\newblock Accessed: 2025-04-16.

\bibitem[{Arias~Jim{\'e}nez et~al.(2022)Arias~Jim{\'e}nez, Rodr{\'\i}guez-Hidalgo, Mier-Sanmart{\'\i}n, and Coronel-Salas}]{arias2022use}
Arias~Jim{\'e}nez, B.; Rodr{\'\i}guez-Hidalgo, C.; Mier-Sanmart{\'\i}n, C.; and Coronel-Salas, G. 2022.
\newblock Use of chatbots for news verification.
\newblock In \emph{Communication and Applied Technologies: Proceedings of ICOMTA 2022}, 133--143. Springer.

\bibitem[{{Ask Solem and Contributors}(2023)}]{celery}
{Ask Solem and Contributors}. 2023.
\newblock Celery - Distributed Task Queue.
\newblock \url{https://docs.celeryq.dev/en/stable/getting-started/introduction.html}.
\newblock Accessed: 2025-04-16.

\bibitem[{Ayers et~al.(2023)Ayers, Poliak, Dredze, Leas, Zhu, Kelley, Faix, Goodman, Longhurst, Hogarth et~al.}]{ayers2023comparing}
Ayers, J.~W.; Poliak, A.; Dredze, M.; Leas, E.~C.; Zhu, Z.; Kelley, J.~B.; Faix, D.~J.; Goodman, A.~M.; Longhurst, C.~A.; Hogarth, M.; et~al. 2023.
\newblock Comparing physician and artificial intelligence chatbot responses to patient questions posted to a public social media forum.
\newblock \emph{JAMA internal medicine}, 183(6): 589--596.

\bibitem[{Bastos and Mercea(2019)}]{bastos2019brexit}
Bastos, M.~T.; and Mercea, D. 2019.
\newblock The Brexit botnet and user-generated hyperpartisan news.
\newblock \emph{Social science computer review}, 37(1): 38--54.

\bibitem[{Calderaro(2018)}]{calderaro2018social}
Calderaro, A. 2018.
\newblock Social media and politics.
\newblock \emph{The SAGE handbook of political sociology}, 2: 781--795.

\bibitem[{{Chirper}(2025)}]{chirper}
{Chirper}. 2025.
\newblock Chirper AI - AI Life Simulation.
\newblock \url{https://chirper.ai/}.
\newblock Accessed: 2025-04-16.

\bibitem[{{Cookiecutter Django}(2025)}]{appstack}
{Cookiecutter Django}. 2025.
\newblock Cookiecutter Django's Documentation.
\newblock \url{https://github.com/cookiecutter/cookiecutter-django}.
\newblock Accessed: 2025-04-16.

\bibitem[{De~Choudhury et~al.(2010)De~Choudhury, Sundaram, John, and Seligmann}]{de2010analyzing}
De~Choudhury, M.; Sundaram, H.; John, A.; and Seligmann, D.~D. 2010.
\newblock Analyzing the dynamics of communication in online social networks.
\newblock \emph{Handbook of social network technologies and applications}, 59--94.

\bibitem[{DiFranzo and Bazarova(2018)}]{difranzo2018truman}
DiFranzo, D.; and Bazarova, N. 2018.
\newblock The Truman Platform: Social Media Simulation for Experimental Research.
\newblock In \emph{ICSWM Workshop" Bridging the Lab and the Field. https://socialmedialab. cornell. edu/the-truman-platform}.

\bibitem[{Felt(2016)}]{felt2016social}
Felt, M. 2016.
\newblock Social media and the social sciences: How researchers employ Big Data analytics.
\newblock \emph{Big data \& society}, 3(1): 2053951716645828.

\bibitem[{Grady(2015)}]{grady2015institutional}
Grady, C. 2015.
\newblock Institutional review boards: Purpose and challenges.
\newblock \emph{Chest}, 148(5): 1148--1155.

\bibitem[{Himelein-Wachowiak et~al.(2021)Himelein-Wachowiak, Giorgi, Devoto, Rahman, Ungar, Schwartz, Epstein, Leggio, and Curtis}]{himelein2021bots}
Himelein-Wachowiak, M.; Giorgi, S.; Devoto, A.; Rahman, M.; Ungar, L.; Schwartz, H.~A.; Epstein, D.~H.; Leggio, L.; and Curtis, B. 2021.
\newblock Bots and misinformation spread on social media: implications for COVID-19.
\newblock \emph{Journal of medical Internet research}, 23(5): e26933.

\bibitem[{Howard and Kollanyi(2016)}]{howard2016bots}
Howard, P.~N.; and Kollanyi, B. 2016.
\newblock Bots,\# strongerin, and\# brexit: Computational propaganda during the uk-eu referendum.
\newblock \emph{arXiv preprint arXiv:1606.06356}.

\bibitem[{Hu et~al.(2025)Hu, Liakopoulos, Wei, Marculescu, and Yadwadkar}]{hu2025simulating}
Hu, T.; Liakopoulos, D.; Wei, X.; Marculescu, R.; and Yadwadkar, N.~J. 2025.
\newblock Simulating Rumor Spreading in Social Networks using LLM Agents.
\newblock \emph{arXiv preprint arXiv:2502.01450}.

\bibitem[{Jiang et~al.(2022)Jiang, Cheng, Yang, and Gao}]{jiang2022ai}
Jiang, H.; Cheng, Y.; Yang, J.; and Gao, S. 2022.
\newblock AI-powered chatbot communication with customers: Dialogic interactions, satisfaction, engagement, and customer behavior.
\newblock \emph{Computers in Human Behavior}, 134: 107329.

\bibitem[{{Kaggle}(2025)}]{kaggle}
{Kaggle}. 2025.
\newblock Kaggle Datasets.
\newblock \url{https://www.kaggle.com/datasets}.
\newblock Accessed: 2025-04-16.

\bibitem[{Kaul et~al.(2015)Kaul, Chaudhri, Cherian, Freberg, Mishra, Kumar, Pridmore, Lee, Rana, Majmudar et~al.}]{kaul2015social}
Kaul, A.; Chaudhri, V.; Cherian, D.; Freberg, K.; Mishra, S.; Kumar, R.; Pridmore, J.; Lee, S.~Y.; Rana, N.; Majmudar, U.; et~al. 2015.
\newblock Social media: The new mantra for managing reputation.
\newblock \emph{Vikalpa}, 40(4): 455--491.

\bibitem[{Krishnan et~al.(2022)Krishnan, Gupta, Gupta, and Singh}]{krishnan2022impact}
Krishnan, C.; Gupta, A.; Gupta, A.; and Singh, G. 2022.
\newblock Impact of artificial intelligence-based chatbots on customer engagement and business growth.
\newblock In \emph{Deep learning for social media data analytics}, 195--210. Springer.

\bibitem[{Kruse, Norris, and Flinchum(2018)}]{kruse2018social}
Kruse, L.~M.; Norris, D.~R.; and Flinchum, J.~R. 2018.
\newblock Social media as a public sphere? Politics on social media.
\newblock \emph{The Sociological Quarterly}, 59(1): 62--84.

\bibitem[{{Lauren Stewart}(2025)}]{social_media}
{Lauren Stewart}. 2025.
\newblock Social Media Research: Analysis of Social Media Data.
\newblock \url{https://atlasti.com/research-hub/social-media-research}.
\newblock Accessed: 2025-04-16.

\bibitem[{Leung and Yan~Chan(2020)}]{leung2020retail}
Leung, C.~H.; and Yan~Chan, W.~T. 2020.
\newblock Retail chatbots: The challenges and opportunities of conversational commerce.
\newblock \emph{Journal of Digital \& Social Media Marketing}, 8(1): 68--84.

\bibitem[{{MCP}(2025)}]{mcp}
{MCP}. 2025.
\newblock Model Context Protocol.
\newblock \url{https://modelcontextprotocol.io/introduction}.
\newblock Accessed: 2025-04-16.

\bibitem[{M{\o}nsted et~al.(2017)M{\o}nsted, Sapie{\.z}y{\'n}ski, Ferrara, and Lehmann}]{monsted2017evidence}
M{\o}nsted, B.; Sapie{\.z}y{\'n}ski, P.; Ferrara, E.; and Lehmann, S. 2017.
\newblock Evidence of complex contagion of information in social media: An experiment using Twitter bots.
\newblock \emph{PloS one}, 12(9): e0184148.

\bibitem[{Nosrati et~al.(2020)Nosrati, Sabzali, Heidari, Sarfi, and Sabbar}]{nosrati2020chatbots}
Nosrati, S.; Sabzali, M.; Heidari, A.; Sarfi, T.; and Sabbar, S. 2020.
\newblock Chatbots, counselling, and discontents of the digital life.
\newblock \emph{Journal of Cyberspace Studies}, 4(2): 153--172.

\bibitem[{Park et~al.(2023)Park, O'Brien, Cai, Morris, Liang, and Bernstein}]{park2023generative}
Park, J.~S.; O'Brien, J.; Cai, C.~J.; Morris, M.~R.; Liang, P.; and Bernstein, M.~S. 2023.
\newblock Generative agents: Interactive simulacra of human behavior.
\newblock In \emph{Proceedings of the 36th annual acm symposium on user interface software and technology}, 1--22.

\bibitem[{{People+AI Research (PAIR) Initiative}(2024)}]{deliberate}
{People+AI Research (PAIR) Initiative}. 2024.
\newblock Deliberate Lab.
\newblock \url{https://github.com/PAIR-code/deliberate-lab}.
\newblock Accessed: 2025-04-16.

\bibitem[{Prabowo et~al.(2008)Prabowo, Thelwall, Hellsten, and Scharnhorst}]{prabowo2008evolving}
Prabowo, R.; Thelwall, M.; Hellsten, I.; and Scharnhorst, A. 2008.
\newblock Evolving debates in online communication: a graph analytical approach.
\newblock \emph{Internet Research}, 18(5): 520--540.

\bibitem[{{Python Software Foundation}(2025{\natexlab{a}})}]{beautiful_soup}
{Python Software Foundation}. 2025{\natexlab{a}}.
\newblock Beautiful Soup.
\newblock \url{https://pypi.org/project/beautifulsoup4/}.
\newblock Accessed: 2025-04-16.

\bibitem[{{Python Software Foundation}(2025{\natexlab{b}})}]{profanity_check}
{Python Software Foundation}. 2025{\natexlab{b}}.
\newblock Profanity Check.
\newblock \url{https://pypi.org/project/profanity-check/}.
\newblock Accessed: 2025-04-16.

\bibitem[{Radivojevic, Clark, and Brenner(2024)}]{radivojevic2024llms}
Radivojevic, K.; Clark, N.; and Brenner, P. 2024.
\newblock Llms among us: Generative ai participating in digital discourse.
\newblock In \emph{Proceedings of the AAAI Symposium Series}, volume~3, 209--218.

\bibitem[{Radivojevic et~al.(2024)Radivojevic, McAleer, Conley, Kennedy, and Brenner}]{radivojevic2024social}
Radivojevic, K.; McAleer, C.; Conley, C.; Kennedy, C.; and Brenner, P. 2024.
\newblock Social Media Bot Policies: Evaluating Passive and Active Enforcement.
\newblock \emph{arXiv preprint arXiv:2409.18931}.

\bibitem[{{Software Freedom Conservancy}(2025)}]{selenium}
{Software Freedom Conservancy}. 2025.
\newblock Selenium.
\newblock \url{https://www.selenium.dev/}.
\newblock Accessed: 2025-04-16.

\bibitem[{Stella, Ferrara, and De~Domenico(2018)}]{stella2018bots}
Stella, M.; Ferrara, E.; and De~Domenico, M. 2018.
\newblock Bots increase exposure to negative and inflammatory content in online social systems.
\newblock \emph{Proceedings of the National Academy of Sciences}, 115(49): 12435--12440.

\bibitem[{Stocking and Sumida(2018)}]{stocking2018social}
Stocking, G.; and Sumida, N. 2018.
\newblock Social media bots draw public’s attention and concern.
\newblock \emph{Pew Research Center}.

\bibitem[{Suarez-Lledo and Alvarez-Galvez(2022)}]{suarez2022assessing}
Suarez-Lledo, V.; and Alvarez-Galvez, J. 2022.
\newblock Assessing the role of social bots during the COVID-19 pandemic: infodemic, disagreement, and criticism.
\newblock \emph{Journal of Medical Internet Research}, 24(8): e36085.

\bibitem[{{Twitter - Developer Platform}(2025)}]{twitter}
{Twitter - Developer Platform}. 2025.
\newblock Twitter API v2: Early Access.
\newblock \url{https://developer.x.com/en/docs/x-api/early-access}.
\newblock Accessed: 2025-04-16.

\bibitem[{Vaidyam et~al.(2019)Vaidyam, Wisniewski, Halamka, Kashavan, and Torous}]{vaidyam2019chatbots}
Vaidyam, A.~N.; Wisniewski, H.; Halamka, J.~D.; Kashavan, M.~S.; and Torous, J.~B. 2019.
\newblock Chatbots and conversational agents in mental health: a review of the psychiatric landscape.
\newblock \emph{The Canadian Journal of Psychiatry}, 64(7): 456--464.

\bibitem[{{Vivian Ho}(2025)}]{zurichreddit}
{Vivian Ho}. 2025.
\newblock Reddit slams ‘unethical experiment’ that deployed secret AI bots in forum.
\newblock \url{https://www.washingtonpost.com/technology/2025/04/30/reddit-ai-bot-university-zurich/}.
\newblock Accessed: 2025-05-15.

\bibitem[{Yang et~al.(2024)Yang, Zhang, Zheng, Jiang, Gan, Wang, Ling, Chen, Ma, Dong et~al.}]{yang2024oasis}
Yang, Z.; Zhang, Z.; Zheng, Z.; Jiang, Y.; Gan, Z.; Wang, Z.; Ling, Z.; Chen, J.; Ma, M.; Dong, B.; et~al. 2024.
\newblock Oasis: Open agents social interaction simulations on one million agents.
\newblock \emph{arXiv preprint arXiv:2411.11581}.

\end{thebibliography}

\end{document}